\documentclass[11pt]{article}
\pdfoutput=1
\usepackage{color}
\usepackage{cite}
\usepackage{graphicx}
\usepackage{multirow}

\newcommand{\be}{\begin{equation}}
\newcommand{\ee}{\end{equation}}
\newcommand{\bea}{\begin{eqnarray}}
\newcommand{\eea}{\end{eqnarray}}
\newcommand{\lsim}{\stackrel{<}{_\sim}}

\definecolor{nicered}{rgb}{1.0,0.0,0.2}

\begin{document}
\begin{center}

\hfill CERN-PH-TH/2012-049, KCL-PH-TH/2012-09, LCTS/2012-05

\vspace{1.5cm}
{\Large\bf Flavour-Changing Decays of a 125~GeV Higgs-like Particle}
\vspace{1.5cm}

{\bf Gianluca Blankenburg}$^{a,b}$, {\bf John Ellis}$^{c,b}$,  {\bf Gino Isidori}$^{d,b}$

\vspace{0.4cm}

{\em $^a$Dip. di Fisica, Univ. di Roma Tre,  Via della Vasca Navale 84, I-00146 Roma, Italy}\\
{\em $^b$CERN, Theory Division, 1211 Geneva 23, Switzerland} \\
{\em $^c$Theoretical Particle Physics and Cosmology Group, Department of Physics, \\ King's College London, London~WC2R~2LS, UK}\\ 
{\em $^d$INFN,  Laboratori Nazionali di Frascati, Via E. Fermi 40, 00044 Frascati, Italy}

\end{center}

\vskip 0.2in
\setlength{\baselineskip}{0.2in}
\centerline{\bf Abstract}
\vskip 0.1in
{\footnotesize
\noindent
The ATLAS and CMS experiments at the LHC have reported the observation of a possible excess
of events corresponding to a new particle $h$ with mass $\sim 125$~GeV that might be the long-sought Higgs boson, 
or something else. Decyphering the nature of this possible signal
will require constraining the couplings of the $h$ and measuring them as accurately as possible. Here we
analyze the indirect constraints on flavour-changing $h$ decays that are provided by limits on low-energy flavour-changing
interactions. We find that indirect limits in the quark sector impose such strong constraints that flavour-changing
$h$ decays to quark-antiquark pairs are unlikely to be observable at the LHC. On the other hand, the upper
limits on lepton-flavour-changing decays are weaker, and the experimental signatures less challenging. In
particular, we find that either ${\mathcal B}(h \to  \tau \bar \mu + \bar \mu \tau)$ or 
${\mathcal B}(h \to  \tau \bar e + \bar e \tau) $ could be ${\cal O}(10)\%$, i.e., comparable to ${\mathcal B}(h \to  \tau^+ \tau^-)$
and potentially observable at the LHC.}

\section{Introduction}

The LHC experiments ATLAS and CMS have excluded the existence of the Higgs boson of the Standard Model (SM)
below 115.5~GeV and between 127 and 600~GeV, and have reported indications of an
apparent excess of events with a mass around $125$~GeV~\cite{ATLAS,CMS}. It is not yet established whether this
excess is due to a new particle nor, if so, whether this new particle resembles closely the
SM Higgs boson. However, we consider this hint to be sufficiently plausible that it is important to
consider all the available constraints on the possible couplings of a new neutral spin-zero particle $h$ with a 
mass around $125$~GeV, with a view to understanding better its nature.

The LHC phenomenology of a SM-like Higgs boson, with mass around
$125$~GeV, is characterized in the first place by six effective couplings: the couplings of $h$ to
${\bar b} b$, $\tau^+ \tau^-$, $\gamma \gamma$, $W^+ W^-$,  $ZZ$ and $gg$. 
ATLAS and CMS are indeed searching for possible decays of any new neutral 
particle in these flavour-conserving final states (except for $gg$,  whose coupling to $h$ is
accessible only through the production mechanism). 
Within the SM, flavour-changing decays of $h$ are expected to
strongly suppressed and well beyond the LHC reach. However, there are alternatives to the SM Higgs interpretation of the 125~GeV
hint, and in some of these cases relatively large flavour-changing couplings become a
significant possibility. This is the case, for example, of the pseudo-dilaton Higgs boson
look-alike discussed in~\cite{pD}, which is quite compatible with the hint observed by ATLAS
and CMS. Flavour-changing decays of $h$ are expected also in the case of a composite Higgs~\cite{Agashe:2009di} 
in models where the Yukawa couplings are functions of the Higgs field~\cite{Giudice:2008uua}
and in several other extensions of the SM with more than one Higgs field
(see, e.g.,~Ref.~\cite{MultiHiggs} and references therein).
It is therefore important to explore the possible existence and the allowed magnitudes of
flavour-changing couplings of a neutral 125~GeV scalar particle $h$, looking for possible deviations
from SM predictions.

In this paper we adopt a phenomenological bottom-up approach, analyzing 
the flavour-changing couplings of the hypothetical $h$ particle allowed by low-energy data.
Several previous studies of this type have been presented in the recent literature, see, 
e.g., \cite{MultiHiggs,Kanemura:2005hr,Paradisi:2005tk,Gabrielli:2011yn,Davidson:2010xv,Goudelis:2011un}.
However, a systematic analysis of both the quark and lepton sectors and their implications for 
the $h$ decays was still missing.
As we will show,  the available experimental constraints 
 on flavour-changing neutral-current (FCNC) interactions
provide strong bounds on many possible quark- and lepton-flavor-changing couplings.
 However, there are instances where relatively large flavour-changing 
$h$ couplings are still allowed by present data, cases in point being the $h{\bar \tau}\mu$ and
$h{\bar \tau}e$ couplings (as already noticed in~\cite{Davidson:2010xv,Goudelis:2011un}). Specifically, we find that current experimental upper limits on
lepton-flavour-violating processes allow the branching ratio
${\mathcal B}(h \to  \tau \bar \mu + \bar \mu \tau) = {\cal O}(10\%)$, and that this can be obtained without 
particular tuning of the effective couplings. It is also possible that 
${\mathcal B}(h \to  \tau \bar e + \bar e \tau) = {\cal O}(10\%)$, though this possibility could 
be realized only at the expense of some fine-tuning of the corresponding couplings and, 
if realized, would forbid a large ${\mathcal B}(h \to  \tau \bar \mu + \bar \mu \tau)$.
The bound on the $\mu e$ modes
are substantially stronger, implying ${\mathcal B}(h \to  \bar \mu e + \bar e \mu)= O(10^{-9})$
in the absence of fine-tuned cancellations.
  
We note that CMS currently reports a 68\% CL range of $0.8^{+1.2}_{-1.3}$ for a possible
$h \to \tau^+ \tau^-$ signal relative to its SM value~\cite{CMS}, and that in the SM ${\mathcal B}(h \to  \tau^+ \tau^-) \sim 6.5\%$
for a SM Higgs boson weighing 125~GeV. It therefore seems that dedicated searches in the 
LHC experiments might already be able to explore flavour-changing leptonic beyond the limits
imposed by searches for lepton-flavour-violating processes.

On the other hand, the indirect upper bounds on possible quark-flavour-violating couplings of
a scalar with mass 125~GeV are much stronger, and the detection of hadronic flavour-changing
decays are much more challenging, so these offer poorer prospects for direct detection at the LHC.

\section{Effective Lagrangian}

We employ here a strictly phenomenological approach, considering the following effective Lagrangian to describe the possible
flavour-changing couplings of a possible neutral scalar boson $h$ to SM quarks and leptons:
\bea
&& \!\!\!
{\mathcal L}_{\rm eff } = 
\sum_{i,j=d,s,b ~ (i\not=j)}  c_{ij} ~\bar d^i_L  d^j_R h +    \sum_{i,j=u,c,t ~ (i\not=j)}  c_{ij} ~\bar u^i_L  u^j_R h  + \sum_{i,j=e,\mu,\tau ~ (i\not=j)}  c_{ij} ~\bar \ell^i_L  \ell^j_R h +  {\rm H.c.}\quad 
\label{eq:effcoupl}
\eea
The field $h$ can be identified with the physical Higgs boson of the SM or, more generally, with a mass eigenstate resulting 
from the mixing of other scalar fields present in the underlying theory with the SM Higgs (if it exists).
Therefore, the operators in (\ref{eq:effcoupl}) are not necessarily $SU(2)_L\times U(1)_Y$ invariant. 
However, they may be regarded as resulting from higher-order $SU(2)_L\times U(1)_Y$-invariant
operators after the spontaneous breaking of $SU(2)_L\times U(1)_Y$. 

By construction, the effective couplings described by (\ref{eq:effcoupl}) are 
momentum-independent.  In principle, higher-order operators with derivative 
couplings could also appear, leading to moment- um-dependent terms,
or effective form factors for the flavour-changing vertices. We assume here
that any such effects are subleading, though it is clear that direct observation
of $h$ decays would, in general, provide much more stringent constraints on
such momentum dependence than could be provided by the indirect low-energy
constraints considered below.

\begin{figure}[t]
\begin{center}
\includegraphics[width=0.15\textwidth]{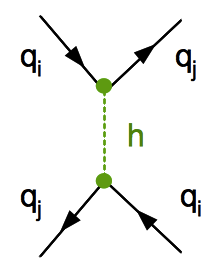}
\hskip 1 cm
\includegraphics[width=0.35\textwidth]{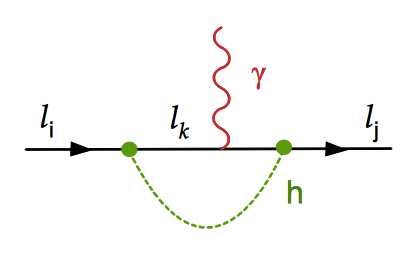}
\vskip -1 cm
\end{center}
\caption{\label{fig:diags} Left: Tree-level diagram contributing to $\Delta F=2$ amplitudes.
Right: One-loop diagram contributing to anomalous magnetic moments and electric dipole
moments of charged leptons ($i=j$), or radiative LFV decay modes  ($i\not=j$). }
\end{figure}

\begin{table}[t]
\begin{center}
\begin{tabular}{c|c|c|c|c} \hline\hline
\rule{0pt}{1.2em}%
Operator &  Eff. couplings &  \multicolumn{2}{c|}{$95\%$ C.L. Bound}  & Observables \\
               &                       &  $|c_{\rm eff} |$  &   $| {\rm Im}(c_{\rm eff}) |$ &  \\ \hline   
($\bar s_R\, d_L)(\bar s_L d_R$)   &  $c_{sd} ~c^*_{ds} $ 
&$1.1 \times 10^{-10}$& $4.1 \times 10^{-13}$  &  $\Delta m_K$; $\epsilon_K$  \\ 
($\bar s_R\, d_L)^2$,~ $(\bar s_L d_R)^2$ &  $c_{ds}^2$,~ $c_{sd}^2 $  
&$2.2 \times 10^{-10}$&  $0.8 \times 10^{-12}$  &   \\ \hline
($\bar c_R\, u_L)(\bar c_L u_R$)   &   $c_{cu} ~c^*_{uc} $  
&$0.9 \times 10^{-9}$& $1.7 \times 10^{-10}$ &  $\Delta m_D$; $|q/p|, \phi_D$  \\  
$(\bar c_R\, u_L)^2$,~ $(\bar c_L u_R)^2$   &  $c_{uc}^2$, ~$c_{cu}^2$ 
& $1.4 \times 10^{-9}$  & $2.5 \times 10^{-10}$  &  \\ \hline
($\bar b_R\, d_L)(\bar b_L d_R)$ &  $c_{bd} ~c_{db}^* $
&  $0.9 \times 10^{-8}$ &   $2.7 \times 10^{-9}$   & $\Delta m_{B_d}$; $S_{B_d \to \psi K}$    \\ 
$(\bar b_R\, d_L)^2$,~$(\bar b_L d_R)^2$  &  $c_{db}^2$,~ $c_{bd}^2 $  
&  $1.0  \times 10^{-8}$ &  $3.0 \times 10^{-9}$ &  \\  \hline
($\bar b_R \,s_L)(\bar b_L s_R)$  &  $c_{bs} ~c_{sb}^* $ 
& $2.0\times10^{-7}$ &  $2.0\times10^{-7}$  & $\Delta m_{B_s}$ \\
$(\bar b_R \,s_L)^2$,~$(\bar b_L s_R)^2$  &  $c_{sb}^2$, ~$c_{bs}^2$ 
& $2.2\times10^{-7}$ & $2.2\times10^{-7}$   &  \\   \hline\hline
\end{tabular}
\caption{\label{tab:DF2} Bounds on combinations of the flavour-changing $h$ couplings 
defined in (\ref{eq:effcoupl}) obtained from $\Delta F=2$ processes~\cite{Isidori:2010kg}, assuming that $m_h = 125$ GeV. }
\end{center}
\end{table}

\section{Bounds  in the Quark Sector}
\label{sect:quarkbounds}

In the quark sector, strong bounds on all the effective couplings in (\ref{eq:effcoupl}) involving light quarks 
(i.e., excluding the top) can be derived from the tree-level contributions to meson-antimeson mixing
induced by diagrams of the type shown in the left panel of Fig.~\ref{fig:diags}.
Using the bounds on dimension-six $\Delta F=2$ operators reported in~\cite{Isidori:2010kg}, 
we derive the indirect limits on different combinations of $c_{ij}$ couplings reported in 
Table~\ref{tab:DF2}. As we discuss in Section~\ref{sect:Higgsdec},
these bounds forbid any flavour-changing decay of the $h$ into a pair of quarks with a
branching ratio exceeding $10^{-3}$. 

The  $\Delta F=1$ bounds on the $c_{ij}$ also prevent sizable Higgs-mediated contributions in 
$\Delta F=1$ amplitudes, {\it if} the flavour-diagonal couplings of the $h$ are the same as the
SM Yukawa couplings. In Table~\ref{tab:DF1} we report the bounds on the 
$c_{ij}$ couplings obtained from $B_{s,d}\to\mu^+\mu^-$ obtained under this assumption, 
namely setting $c_{\mu\mu}=\sqrt{2} m_\mu/v$ with $v\approx 246$~GeV~\footnote{~This assumption 
is not true in general. For example, in the
pseudo-dilaton scenario of~\cite{pD} the flavour-diagonal $h$ couplings are in general
suppressed by a universal factor $c < 1$, in which case the bounds in Table~\ref{tab:DF1}
would be weakened by a factor $1/c > 1$.}. As can be seen, 
these $\Delta F=1$ bounds are weaker than those in Table~\ref{tab:DF2}.
This would not be true if the flavour-diagonal couplings of $h$ were enhanced with respect to 
the SM Yukawa couplings, or if there were some extra contribution cancelling $h$-exchange 
in the $\Delta F=2$ amplitudes. The latter happens, for instance,  in some two-Higgs doublet models,
because of the destructive interference of scalar and pseudo-scalar exchange amplitudes:
see, e.g., \cite{MultiHiggs,Gerard}.

\begin{table}[t]
\begin{center}
\begin{tabular}{c|c|c} \hline\hline
\rule{0pt}{1.2em}%
 Eff. couplings &  Bound   & Constraint \\  \hline              
                  $|c_{sb}|^2$,~ $|c_{bs}|^2$  &   $2.9  \times 10^{-5}$~[*]     &  ${\mathcal B}(B_s\to\mu^+\mu^-) < 1.4 \times 10^{-8}$   \\ 
                  $|c_{db}|^2$,~ $|c_{bd}|^2$    &    $1.3  \times 10^{-5}$~[*]    &  ${\mathcal B}(B_d\to\mu^+\mu^-) < 3.2 \times 10^{-9} $    \\  \hline\hline
\end{tabular}
\caption{\label{tab:DF1} Bounds on combinations of the flavour-changing $h$ couplings 
defined in (\ref{eq:effcoupl}) obtained from experimental constraints on rare $B$ decays~\cite{LHCb}, assuming that $m_h = 125$ GeV.
(Here and in subsequent Tables, the [*] denotes bounds  obtained under the assumption that the flavour-diagonal 
couplings of $h$ are the same as the corresponding SM Yukawa couplings.)}
\end{center}
\end{table}

\begin{table}[t]
\begin{center}
\begin{tabular}{c|c|c|c} \hline\hline
\rule{0pt}{1.2em}%
Operator&  Eff. couplings &   Bound & Constraint  \\  \hline
                $(\bar \mu_R\, e_L)(\bar q_L q_R)$,~ $(\bar \mu_L\,  e_R)(\bar q_L q_R)$
                 &  $|c_{\mu e}|^2$,~ $|c_{e \mu}|^2$  &    $3.0\times  10^{-8}$~[*] &  ${\mathcal B}_{\mu \to e}({\rm Ti})< 4.3 \times 10^{-12}$        \\  \hline
               $(\bar \tau_R\, \mu_L)(\bar \mu_L \mu_R)$,~ $(\bar \tau_L\, \mu_R)(\bar \mu_L \mu_R)$    
                 &  $|c_{\tau \mu}|^2$,~ $|c_{\mu \tau}|^2$ &   $2.0 \times 10^{-1}$~[*]     &  $\Gamma(\tau \to\mu \bar \mu \mu ) 	<2.1 \times  10^{-8} $       \\ 
               $(\bar \tau_R\, e_L)(\bar \mu_L \mu_R)$,~ $(\bar \tau_L\,  e_R)(\bar \mu_L \mu_R)$
                 &  $|c_{\tau e}|^2$,~ $|c_{e \tau}|^2$    &    $4.8 \times 10^{-1}$~[*]   &  $\Gamma(\tau \to e \bar \mu \mu ) 	<2.7  \times  10^{-8}$    \\ \hline 
               $(\bar \tau_R\, e_L)(\bar \mu_L e_R)$,~ $(\bar \tau_L\, e_R)(\bar \mu_L  e_R)$    
                 &   $|c_{\mu e } c_{e\tau}^*|$,~  $|c_{\mu e } c_{\tau e}|$  &    $0.9 \times 10^{-4}$  &  $\Gamma(\tau \to \bar \mu   e e )	<1.5 \times  10^{-8}$         \\ 
                  $(\bar \tau_R\, e_L)(\bar  \mu_R e_L)$,~ $(\bar \tau_L\, e_R)(\bar \mu_R   e_L  )$    
                 &  $|c_{e \mu}^* c_{e\tau}^*|$,~  $|c_{e \mu}^* c_{\tau e}|$  &       &    \\ \hline
                $(\bar \tau_R\,  \mu_L)(\bar  e_L \mu_R)$,~ $(\bar \tau_L\,  \mu_R)(\bar e_L \mu_R)$
                 &   $|c_{e \mu} c_{\mu \tau }^*|$,~ $|c_{e \mu} c_{\tau \mu}|$   &     $1.0 \times 10^{-4}$    &  $\Gamma(\tau \to  \bar e \mu \mu )<1.7 \times  10^{-8}$     \\
                $(\bar \tau_R\,  \mu_L)(\bar   e_R \mu_L)$,~ $(\bar \tau_L\,  \mu_R)(\bar e_R \mu_L )$
                 &  $|c_{\mu e}^* c_{\mu \tau }^*|$,~ $|c_{\mu e}^* c_{\tau \mu}|$     &        &    \\ \hline \hline
\end{tabular}
\caption{\label{tab:LFVbounds1} Bounds on combinations of the flavour-changing $h$ couplings 
defined in (\ref{eq:effcoupl}) obtained from  charged-lepton-flavour-violating decays, assuming that $m_h = 125$ GeV.
}
\end{center}
\end{table}

\section{Bounds in the Lepton Sector}
\label{sect:leptonbounds}

In the lepton sector we do not have an analogous of the $\Delta F=2$ constraints, leaving more room
for sizeable non-standard contributions. 

We start by analyzing the tree-level contributions of $h$ 
to the lepton-flavour violating (LFV) decays of charged leptons and $\mu\to e$ conversion in nuclei. In most cases
bounds on the effective couplings in (\ref{eq:effcoupl}) 
 can be derived only with an Ansatz about the flavour-diagonal couplings. Here we
 assume again that the flavour-diagonal couplings are the SM Yukawas,
 \be
 c_{\ell \ell} = y_\ell \equiv \frac{\sqrt{2} m_\ell}{v}~.
 \ee
This leads to the bounds reported in Table~\ref{tab:LFVbounds1}, where we have used the limits of the 
corresponding dimension-six operators reported in~\cite{Raidal:2008jk}, updating the results on various 
$\tau$ decay modes from Ref.~\cite{PDG}.
As can be  seen, all the bounds except that derived from $\mu\to e$ conversion\footnote{~The bound from 
$\mu\to e$ conversion  has been derived following the recent analysis of Ref.~\cite{Cirigliano:2009bz}:
the dominant 
constraint follows from ${\mathcal B}_{\mu \to e}({\rm Ti})$ and, in order to derive a 
conservative bound, we have set $ y =  2\langle N | \bar s s | N\rangle /\langle N | \bar d d+ \bar u u | N\rangle=0.03$.} 
are  quite weak.\footnote{~As commented previously, in the scenario of Ref.~\cite{pD} the flavour-diagonal $h$ couplings are in general
suppressed by a universal factor $c < 1$, in which case the first three bounds in Table~\ref{tab:LFVbounds1}
would be  weakened by a factor $1/c > 1$.} Note in particular that if we impose 
$c_{\mu e},c_{e\mu} < y_\mu \approx 6 \times 10^{-4}$ we have essentially no bounds on the 
flavour-violating couplings involving the $\tau$ lepton. Note also that we cannot 
profit from the strong experimental bound on  $\Gamma(\mu \to   e \bar e e)$, since the corresponding amplitude is
strongly suppressed by the electron Yukawa coupling. 

Next we proceed to analyze  one-loop-induced amplitudes. 
At the one-loop level the flavour-violating couplings in (\ref{eq:effcoupl}) induce: (i) logarithmically-divergent 
corrections to the lepton masses;  (ii) finite contributions to the anomalous magnetic moments and the electric-dipole moments (edms)
of charged leptons; and (iii) finite contributions to radiative LFV decays of the type 
$l_i \to l_j \gamma$ (see the right panel of Fig.~\ref{fig:diags}).

\begin{table}[t]
\begin{center}
\begin{tabular}{c|c|c} \hline\hline
\rule{0pt}{1.2em}%
  Eff. couplings &  Bound  & Constraint  \\  \hline
                $|c_{e\tau}c_{\tau e}| \quad \left( |c_{e\mu}c_{\mu e}| \right)$ &    $1.1 \times 10^{-2}\quad (1.8\times 10^{-1})$  &  $|\delta m_e| < m_e$   \\  
               $|{\rm Re}(c_{e\tau}c_{\tau e})|  \quad \left( |{\rm Re}(c_{e\mu}c_{\mu e})| \right)$   &    $ 0.6 \times 10^{-3}\quad (0.6  \times 10^{-2})$ &  $|\delta a_e| <  6 \times 10^{-12}$   \\  
               $|{\rm Im}(c_{e\tau}c_{\tau e})|  \quad \left( |{\rm Im}(c_{e\mu}c_{\mu e})| \right)$    &    $ 0.8 \times 10^{-8}\quad (0.8  \times 10^{-7})$   &  $ |d_e| <  1.6\times 10^{-27}~e{\rm cm}$   \\    
               $|c_{\mu\tau}c_{\tau \mu}| $       &   $ 2 $  &   $|\delta m_\mu| < m_\mu$  \\  
               $|{\rm Re}(c_{\mu\tau}c_{\tau \mu})| $     &   $  2\times 10^{-3}$   &  $|\delta a_\mu| <   4 \times 10^{-9}$   \\  
               $|{\rm Im}(c_{\mu\tau}c_{\tau \mu})|  $      &    $ 0.6 $  &  $|d_\mu| <   1.2\times 10^{-19}~e{\rm cm}$    \\   
               $|c_{e\tau}c_{\tau\mu}|, ~ |c_{\tau e}c_{\mu\tau}|$		
               &	$1.7\times 10^{-7}$	&	$\mathcal{B}(\mu\rightarrow e\gamma)< 2.4\times 10^{-12}$  \\
               $|c_{\mu\tau}|^2, ~ |c_{\tau\mu}|^2$	&	$0.9\times 10^{-2}$~[*]  &	$\mathcal{B}(\tau\rightarrow \mu\gamma)< 4.4\times 10^{-8}$		 \\
               $|c_{e\tau}|^2, ~ |c_{\tau e}|^2$		&		$0.6\times 10^{-2}$~[*]  &	$\mathcal{B}(\tau\rightarrow e\gamma)< 3.3\times 10^{-8}$ \\
               \hline \hline
              \end{tabular}
\caption{\label{tab:mass} Bounds on combinations of the flavour-changing $h$ couplings 
defined in (\ref{eq:effcoupl}) obtained from  the naturalness requirement $|\delta m_\ell | < m_\ell$
(assuming  $\Lambda = 1$ TeV), from the contributions to $a_\ell$ and $d_\ell$ ($\ell=e,\mu$),
and from radiative LFV decays (in all cases we set $m_h = 125$ GeV.
}
\end{center}
\end{table}

As far as the mass corrections are concerned, in the leading-logarithmic approximation we find\footnote{~The complex mass
correction $\delta m_\ell$ is defined by $m_\ell \bar \ell \ell   \longrightarrow   \bar \ell  \left[  m_\ell  + {\rm Re}(\delta m_\ell) + i {\rm Im}(\delta m_\ell) \gamma_5 \right] \ell $.}
\be
\delta m_\ell  =  \frac{1}{(4\pi)^2}  \sum_{j\not=\ell}  c_{\ell j} c_{j \ell }  
m_j\log\left(\frac{m_h^2}{\Lambda ^2}\right)~.
\ee
In absence of fine-tuning we expect $|\delta m_\ell | < m_\ell$ for each of the two possible contributions in the sum. The most significant 
bounds thus derived, setting $\Lambda =1$~TeV, are reported in Table~\ref{tab:mass}. Note that in this case no 
assumption on the flavour-diagonal couplings is needed. 

More stringent (and more physical) bounds on the same combinations 
of couplings are derived from the contributions to the anomalous magnetic moments, 
$a_\ell = (g_\ell -2)/2$ and the edms of the electron and the muon.
The corresponding one-loop amplitudes are
\bea
|\delta a_\ell | &=& \frac{ 4 m_\ell^2}{m_h^2} \frac{1}{(4\pi)^2}  \sum_{j\not=\ell}  {\rm Re} (  c_{\ell j} c_{j \ell }  ) \frac{m_j}{m_\ell }\left(\log\frac{m_h^2}{m_\tau^2}-\frac{3}{2}\right)~, \\
|d_\ell | &=&   \frac{ 2 m_\ell}{m_h^2}  \frac{e}{(4\pi)^2}  \sum_{j\not=\ell}  {\rm Im} (  c_{\ell j} c_{j \ell }  ) \frac{m_j}{m_\ell}\left(\log\frac{m_h^2}{m_\tau^2}-\frac{3}{2}\right)~, 
\eea
from which we derive the bounds reported in Table~\ref{tab:mass}.\footnote{~As usual, 
we define $a_\ell$ and $d_\ell$ in terms of the couplings of the corresponding dipole operators as follows:
$( e a_\ell/4 m_\ell) \bar \ell  \sigma_{\mu\nu}  \ell F^{\mu\nu}$,  $i(d_\ell/2) \bar \ell \sigma_{\mu\nu} \gamma_ 5 \ell F^{\mu\nu} $.
The error on $\delta a_e$ reported in Table~\ref{tab:mass} is the theoretical error in predicting $(g-2)_e$ using independent 
 determinations of $\alpha_{\rm em}$~\cite{Hanneke:2008tm}. }
We do not report the corresponding bounds from $a_\tau$ and $d_\tau$ since they are much weaker.
As can be seen, with the exception of the bound from the electron edm, which can easily be evaded assuming real couplings,
the bounds are still rather weak. 

The radiative LFV decay rates generated at one loop level can be written as
\begin{equation}
\Gamma(l_i\rightarrow l_j \gamma)=m_i^3\frac{e^2}{16\pi}(|A^L_{ij}|^2+|A^R_{ij}|^2)
\end{equation}
with  coefficients 
\begin{eqnarray}
&&|A^{R}_{\mu e}|= \frac{1}{(4\pi)^2}|c_{e \tau}c_{\tau \mu}| \frac{m_\tau}{m_h^2} \left(\log\frac{m_h^2}{m_\tau^2}-\frac{3}{2}\right),\quad 
|A^{L}_{\mu e}| = \frac{1}{(4\pi)^2}|c_{\tau e }c_{\mu \tau}| \frac{m_\tau}{m_h^2} \left(\log\frac{m_h^2}{m_\tau^2}-\frac{3}{2}\right),\quad \\
&&|A^{R}_{\tau \ell}| =  \frac{1}{(4\pi)^2}|c_{\ell \tau}|y_\tau \frac{m_\tau}{m_h^2} \left(  \log\frac{m_h^2}{m_\tau^2} - \frac{4}{3} \right),\quad 
|A^{L}_{\tau \ell}| =  \frac{1}{(4\pi)^2}|c_{\tau \ell}|y_\tau \frac{m_\tau}{m_h^2}  \left(  \log\frac{m_h^2}{m_\tau^2} - \frac{4}{3} \right),
\end{eqnarray}
and corresponding bounds reported in Table~\ref{tab:mass}.
Here it should be noted the strong and model-independent bound from $\mu\to e\gamma$~\cite{MEG}
which prevents the $h{\bar \tau}\mu$ ($h{\bar \mu}\tau$)  and
$h{\bar \tau}e$ ($h\bar e \tau$) couplings to be both large at the same time.

Finally we consider the bounds coming from two-loop diagrams of Barr-Zee type~\cite{BZ}, with a top-quark loop,
whose relevance in constraining Higgs LFV couplings has been stressed recently in~\cite{Davidson:2010xv,Goudelis:2011un}.
Despite being suppressed by an extra $1/(16\pi^2)$ factor, these amplitudes are proportional to a single 
lepton Yukawa coupling and cannot be neglected.  
The resulting bounds, shown in Table~\ref{tab:2loops}, are obtained under the assumption that the coupling of $h$ to the top quark is the same as in the
SM ($c_{yy} = y_t \equiv  \sqrt{2} m_t/v$). These bounds are consistent with those 
reported in~Ref.~\cite{Goudelis:2011un}.

\begin{table}[t]
\begin{center}
\begin{tabular}{c|c|c} \hline\hline
\rule{0pt}{1.2em}%
  Eff. couplings &  Bound  & Constraint  \\  \hline
               
               $|c_{e\mu}|^2, ~ |c_{\mu e}|^2$		
               &	$1\times 10^{-11}$~[*]	&	$\mathcal{B}(\mu\rightarrow e\gamma)< 2.4\times 10^{-12}$  \\
               $|c_{\mu\tau}|^2, ~ |c_{\tau\mu}|^2$	&	$5\times 10^{-4}$~[*]  &	$\mathcal{B}(\tau\rightarrow \mu\gamma)< 4.4\times 10^{-8}$		 \\
               $|c_{e\tau}|^2, ~ |c_{\tau e}|^2$		&		$3\times 10^{-4}$~[*]  &	$\mathcal{B}(\tau\rightarrow e\gamma)< 3.3\times 10^{-8}$ \\
               \hline \hline
              \end{tabular}
\caption{\label{tab:2loops} Bounds from two-loop Barr-Zee diagrams~\cite{BZ} contributing to LFV decays.
}
\end{center}
\end{table}

\section{Higgs decays}
  \label{sect:Higgsdec}

Normalizing the flavour-violating $h$ decays to the $h \to \tau  \bar \tau$ mode, which we assume to be SM-like, we can write 
\begin{eqnarray}
\frac{{\mathcal B}(h \to f_i  \bar f_j  )}{{\mathcal B}(h \to \tau  \bar \tau )}    \approx   N_f  \times \frac{|c_{ij}|^2+|c_{ji}|^2}{2 y_\tau^2} = 0.48 \times 10^4
\times N_f  \left( |c_{ij}|^2+|c_{ji}|^2 \right) ~,
\end{eqnarray}
where $N_q=3$ and $N_\ell=1$, and we have neglected tiny  $m_{f_{i,j}}/m_h$ corrections.
Assuming ${\mathcal B}(h \to \tau  \bar \tau ) \approx 6.5\%$,  as expected for a SM Higgs boson with $m_h=125$~GeV, we get 
\begin{eqnarray}
{\mathcal B}(h \to f_i  \bar f_j )   \approx  3.1 \times 10^2 \times N_f  \left( |c_{ij}|^2+|c_{ji}|^2 \right) ~.
\end{eqnarray}

In the quark sector, in the most favourable case we get ${\mathcal B}(h \to b  \bar s ,~ \bar s b)  <  4 \times 10^{-4}$, 
which is beyond the reach of the LHC, also in view 
of the difficult experimental signature. However, the situation is much more favourable in the lepton sector. 
From the compilation of bounds in the previous section we derive the following conclusions:
\begin{itemize}
\item ${\mathcal B}(h \to  \tau \bar \mu + \bar \mu \tau) = {\cal O}(10\%)$ does not contradict any experimental bound and
does not require off-diagonal couplings larger 
than the corresponding diagonal ones ($|c_{\mu\tau}|, |c_{\tau\mu}|  \lsim y_\tau$). 
It can be obtained even assuming  ${\cal O}(1)$ CP-violating phases for the $c_{\mu\tau(\tau\mu)}$
couplings, provided  $|c_{e\tau (\tau e)}/c_{\mu\tau (\tau \mu)} |< 10^{-2}$ in order to satisfy the 
$\mu\to e \gamma$ bound.

\item ${\mathcal B}(h \to  \tau \bar e + \bar e \tau) $ can also reach ${\cal O}(10\%)$ values, but only at the price of some tuning of the 
corresponding effective couplings. In particular, 
negligible CP-violating phases are needed in order to satisfy the tight constraint provided by the electron edm shown in Tab.~\ref{tab:mass}. 
Moreover, $|c_{\mu\tau (\tau \mu)}/c_{e \tau  (\tau e)} |<  10^{-2}$ in order to satisfy the  $\mu\to e \gamma$ bound.

\item The $\mu\to e \gamma$ bound implies that only one of ${\mathcal B}(h \to  \tau \bar \mu + \bar \mu \tau)$
or ${\mathcal B}(h \to  \tau \bar e + \bar e \tau)$ could be ${\cal O}(10\%)$.

\item The bounds from $\mu\to e$ conversion in nuclei and from $\mu\to e \gamma$
forbid large branching ratios for the clean $\mu e$ modes. Specifically,
we find ${\mathcal B}(h \to  \bar \mu e + e\bar \mu) < 3\times 10^{-9}$, several orders of magnitude below the
flavour-conserving   ${\mathcal B}(h \to  \mu \bar \mu)  \approx 2.3 \times 10^{-4}$ expected for a 125~GeV SM Higgs.
However, we recall that this strong bound holds under the hypothesis of SM-like flavour-diagonal couplings for $h$.
\end{itemize}

\section{Summary}

The possible observation of a new particle $h$ with mass around 125~GeV raises the important question
of its possible nature: is it a SM-like Higgs boson, or not? Key answers to this question will be provided by
measurements of the $h$ couplings, and ATLAS and CMS have already provided valuable information~\cite{ATLAS,CMS} on
its flavour-diagonal couplings (if the $h$ exists). Further information could be provided by searches for
(and measurements of) its flavour-changing couplings. In this paper we have analyzed the indirect upper bounds
on these couplings that are provided by constraints on flavour-changing and other interactions
in both the quark and lepton sectors.

We have found that in the quark sector the indirect constraints are so strong, and the experimental
possibilities at the LHC so challenging, that quark flavour-changing decays of the $h$ are unlikely to
be observable.

However, the situation is very different in the lepton sector. Here the indirect constraints are typically 
much weaker, and the experimental possibilities much less challenging. Specifically, we find that
either  ${\mathcal B}(h \to  \tau \bar \mu + \bar \mu \tau)$ or ${\mathcal B}(h \to  \tau \bar e + \bar e \tau)$
of order $10\%$ is a  possibility allowed by the available LFV constraints. These 
large partial decay rates are the combined result not only of relatively weak bounds on Higgs-mediated LFV amplitudes involving 
the $\tau$ lepton, but also of the smallness of the total $h$ decay width 
for $m_h \approx 125$~GeV. Interestingly, 
these potentially large LFV rates  
are comparable to the expected branching ratio
for $h \to \tau^+ \tau^-$ in the SM, which is already close to the sensitivity of the CMS experiment~\cite{CMS}.
Therefore the LHC experiments may soon be able to provide complementary information on the LFV
couplings of the (hypothetical) $h$ particle with mass 125~GeV. The decays $h \to  \bar \mu e,  \bar e \mu$
are constrained to have very small branching ratios, but their experimental signatures are so clean that here
also the LHC may soon be able to provide interesting information.

We therefore urge our experimental colleagues to make dedicated searches for these interesting
flavour-violating decays of the possible $h$ particle with mass 125~GeV.

\subsection*{Acknowledgments}
We thank S.~Davidson and O.~Lebedev for useful comments.
This work was supported by the EU ERC Advanced Grant FLAVOUR (267104),
and by MIUR under contract 2008XM9HLM.
The work of J.E. is supported partly by the London
Centre for Terauniverse Studies (LCTS), using funding from the European
Research Council 
via the Advanced Investigator Grant 267352.
G.I. acknowledges the support of the Technische Universit\"at M\"unchen -- Institute for Advanced
Study, funded by the German Excellence Initiative.

\end{document}